\documentclass[aps,twocolumn,prl,showpacs,color,psfig,epsf]{revtex4}
\usepackage{amsmath}
\usepackage{color}
\usepackage{amsfonts}
\usepackage{epsf}
\usepackage{graphicx}
\usepackage{epstopdf}
\begin{document}

\title{Enhanced diffusion of nonswimmers in a three-dimensional bath of motile bacteria}

\author{Alys Jepson, Vincent A. Martinez, Jana Schwarz-Linek, Alexander Morozov, Wilson C. K. Poon}
\affiliation{$^1$SUPA, School of Physics $\&$ Astronomy, The University of Edinburgh, Mayfield Road, Edinburgh EH9 3JZ, UK}

\begin{abstract}
We show using differential dynamic microscopy that the diffusivity of non-motile cells in a 3D population of motile {\it E. coli} is enhanced by an amount proportional to the active cell flux. While non-motile mutants without flagella and mutants with paralysed flagella have quite different thermal diffusivities and therefore hydrodynamic radii, their diffusivities are enhanced to the same extent by swimmers in the regime of cell densities explored here. Integrating the advective motion of non-swimmers caused by swimmers with finite persistence-length trajectories predicts our observations to within 2\%, indicating that fluid entrainment is not relevant for diffusion enhancement in 3D.

\end{abstract}

\maketitle

A collection of swimmers in a liquid (fish, motile algae, Janus colloids in `fuel', \ldots) is an example of intrinsically non-equilibrium `active matter' \cite{Ramaswamy2010Mechanics}, which show multiple intriguing activity-driven phenomena, e.g. novel pattern formation and counter-intuitive rheology \cite{CatesReview2012}. In particular, swimmers perturb the motion of passive species in their vicinity, from turning micro-gear wheels  \cite{Sokolov2010Swimming,Angelani2009SelfStarting} to enhancing the motion of tracer colloids \cite{Wu2000Particle,Mino2010Enhanced,Mino2012Induced,Valeriani2011Colloids,Chen2007Fluctuations,Leptos2009Dynamics}. Understanding such phenomena is challenge to statistical physics; it is also relevant biologically. Motile microorganisms live in the presence of and interact with non-swimmers of the same or different species, and non-living debris that as food, substrates for colonisation, etc. Such active-passive interactions are important ecologically \cite{Grossart2001Bacterial}, e.g. in cross-species predator-prey relationships. 

The most well studied active-passive mixture to date is colloids in a bacterial bath
\cite{Wu2000Particle,Mino2010Enhanced,Mino2012Induced,Valeriani2011Colloids,Chen2007Fluctuations}. Experiments show that swimming bacteria enhance the long-time (non-thermal) diffusivity, $D$, of colloidal tracers linearly with the swimmer concentration \cite{Wu2000Particle}, or, more generally, the active particle flux \cite{Mino2010Enhanced,Mino2012Induced}, $J_A=\bar v n_A$, where $\bar v$ and $n_A$ are the average speed and number density of the swimmers, i.e. 
\begin{equation}
\Delta D = D - D_0 = \beta J_A, \label{beta}
\end{equation} 
with $D_0$ the (thermal) diffusivity in absence of swimmers.

Significantly, all experiments supporting Eq.~\ref{beta} \cite{Wu2000Particle,Mino2010Enhanced,Mino2012Induced,Valeriani2011Colloids} have been  in 2D, with the swimmers in a thin film \cite{Wu2000Particle} or close to one  \cite{Mino2010Enhanced,Mino2012Induced} or two \cite{Mino2012Induced,Valeriani2011Colloids} walls; in \cite{Wu2000Particle}, the swimmers were at interacting concentrations. It remains unknown whether Eq.~\ref{beta} holds under much simpler, bulk (3D) conditions far from any boundaries at low swimmer concentrations. Moreover, existing calculations  \cite{Mino2012Induced} considering only far-field advection of tracer motion \cite{Yeomans2010, Childress2011} significantly underestimate 2D observations \cite{Mino2010Enhanced,Mino2012Induced}. One factor may be the presence of a range of swimmer-wall distances in the experiments. It has also been proposed recently \cite{PushkinPreprint} that advection and fluid entrainment \cite{Pushkin2013}  both contribute in 3D, but entrainment dominates in 2D. This new theory predicts a value of $\beta$ in 3D that is more than an order of magnitude larger than that given in \cite{Mino2012Induced}.

\begingroup
\squeezetable\begin{table}
\caption{Experiments (E) and theory (T)  on tracer diffusion}
\begin{tabular}{|l|l|l|l|}
\hline
 & Dimensionality & Effect(s) included & $\beta$ ($\mu\mbox{m}^4$)\footnotemark \footnotetext{See Eq.~\ref{beta} for the definition of $\beta$; cf. Fig.~\ref{Fig:D}.} \\
\hline
E \cite{Mino2012Induced} & Next to wall &  & $13 \pm 0.7$\footnotemark \footnotetext{Fitted value reported in \protect\cite{Mino2012Induced} based on their Fig.~8.} \\
\hline
T \cite{Mino2012Induced} & Next to wall & Advection & 2.0 \\
\hline
T \cite{Mino2012Induced}  & 3D & Advection & 0.48\footnotemark \footnotetext{Approximating swimmers as point dipoles, as in this work.} \\
\hline
T \cite{PushkinPreprint} & 3D & Advection + Entrainment & 9.0 \\
\hline \hline 
E [this work] & 3D  &  & $7.1 \pm 0.4$ \\
\hline
T [this work] & 3D & Advection & 7.24\\
\hline
\end{tabular}
\end{table}
\endgroup

Thus, the current situation, Table I, is far from satisfactory. To progress, confrontation of theory with 3D data is essential. 
We report a 3D study of enhanced diffusion in a bacterial bath using differential dynamic microscopy (DDM), which is uniquely able to deliver high-throughput 3D averaging  \cite{Cerbino}. We predict the measured $\beta$ to within 2\% by considering advection alone, showing that entrainment is negligible in 3D. 

DDM measures the intermediate scattering function (ISF), $f(q,t)$, of a population swimming {\it E.~coli} \cite{Wilson2011Differential}, where $q$ is the scattering vector and $t$ is time. Fitting the ISF gives the swimming speed distribution, and hence the average speed $\bar v$, the fraction of non-motile organisms, $\alpha$, and the diffusivity of the non-motile species, $D$. The method has been validated in detail for wild-type (WT), i.e.~run-and-tumble, and smooth swimming {\it E.~coli} \cite{Martinez2012Differential}. 

We use non-swimming cells as tracers. Since the fraction of motile organisms in as-prepared (`native') populations do not vary significantly from day to day, we add non-swimmers to native populations to study $\Delta D$ as a function of $J_A$. Thus, in general, there are 3 sub-populations in each of our samples: {\it native} motile (M) and non-motile (N1) cells, and {\it added} non-motile (N2) cells, the latter being fluorescent, and therefore distinguishable from native non-motile cells. 
We performed DDM in phase contrast and fluorescence \cite{LuDDM} modes, probing the motion of all the cells and only the diffusion of the added, fluorescent non-motile mutants (N2) respectively.

K12-derived wild-type (WT) \emph{E. coli} AB1157 and fluorescent non-motile fliF (no flagella) or motA (paralyzed flagella) mutants \cite{mutants}  were grown and harvested as described before \cite{Martinez2012Differential}.
Suspensions at optical density OD = 0.5 (at 600 nm), corresponding to $7.8 \pm 0.2\times10^{8}$ cells/ml (= cell body volume fraction $\phi \approx 0.1\%$ based on cell volume of $V = 1.4 \pm 0.1 \mu\mbox{m}^3$ \cite{cell_volume}), were obtained by dilution. 

DDM showed that as-prepared WT populations (M + N1) contained 20-40\% native non-motile (N1) cells (i.e. $\alpha = 0.6$-0.8), and motile cells swam with $\bar v = 13 -16\mu$m/s \cite{Wilson2011Differential,Martinez2012Differential}. We studied the effect of $J_A = \bar v n_A  = \bar v \alpha\phi/V$ on enhanced diffusion using three protocols. In most cases, we varied $\alpha$ directly by mixing WT and mutant cell suspensions at different ratios to obtain samples with fixed $\phi = 0.1\%$ and $\bar v$ in the narrow range $\bar v = 13 -16\mu$m/s. To check that it is the combination $\bar v\alpha\phi$ that controls $\Delta D$, we repeated these experiments but added glucose ($0.006\text{wt}\%$) into cell mixtures immediately before loading into capillaries, which increased $\bar v$ to $\lesssim 25 \mu$m/s \cite{Adler1967Effect}. Finally, we studied a limited number of mixtures in which we varied $\phi$ at fixed $\alpha$ or varied $\phi$ and $\alpha$ together. Taken together, these experiments accessed $0 \leq J_A \leq 14 \times 10^{-3} \mu$m$^{-2}$s$^{-1}$ by varying the component parameters of $J_A$ in the range $0 \leq \alpha \lesssim 0.7$ and $0.04\% \lesssim \phi \lesssim 0.1 \%$, and $\bar v \approx 15\mu$m/s and $\approx 25\mu$m/s.

Observations began immediately after a glass capillary (depth $400$ $\mu$m) was filled with $\approx200$ $\mu$l of solution and sealed with Vaseline to prevent drift. Forty-second phase-contrast movies (Nikon Plan Fluor 10$\times$ objective, NA = 0.3, 100 frame per second, $500^2$ pixels) capturing all cells  ($\sim 10^4$ M+N1+N2) and fluorescence movies (Nikon Plan Fluor 20$\times$ objective with NA = 0.5, 20~fps, $1024^2$ pixels excited at $450-490$ nm) capturing only the added non-motile mutants  ($\sim 10^2$-$10^3$ N2) were consecutively recorded on an inverted microscope (Nikon TE300 Eclipse) with a Mikrotron high-speed camera (MC 1362) and frame grabber (Inspecta 5, 1Gb memory). We image at $100$ $\mu$m  from the bottom of the capillary. This is significantly larger than the persistence length of WT {\it E. coli} (1s run time $\equiv \,\lesssim 15-20 \mu$m run length), so that they execute 3D motion. We have previously shown that the depth at 10$\times$ or 20$\times$ is large enough for DDM to return the 3D ISF of swimming {\it E. coli} \cite{Martinez2012Differential}.

\begin{figure}
	\begin{center}
	\includegraphics[width=3.4in]{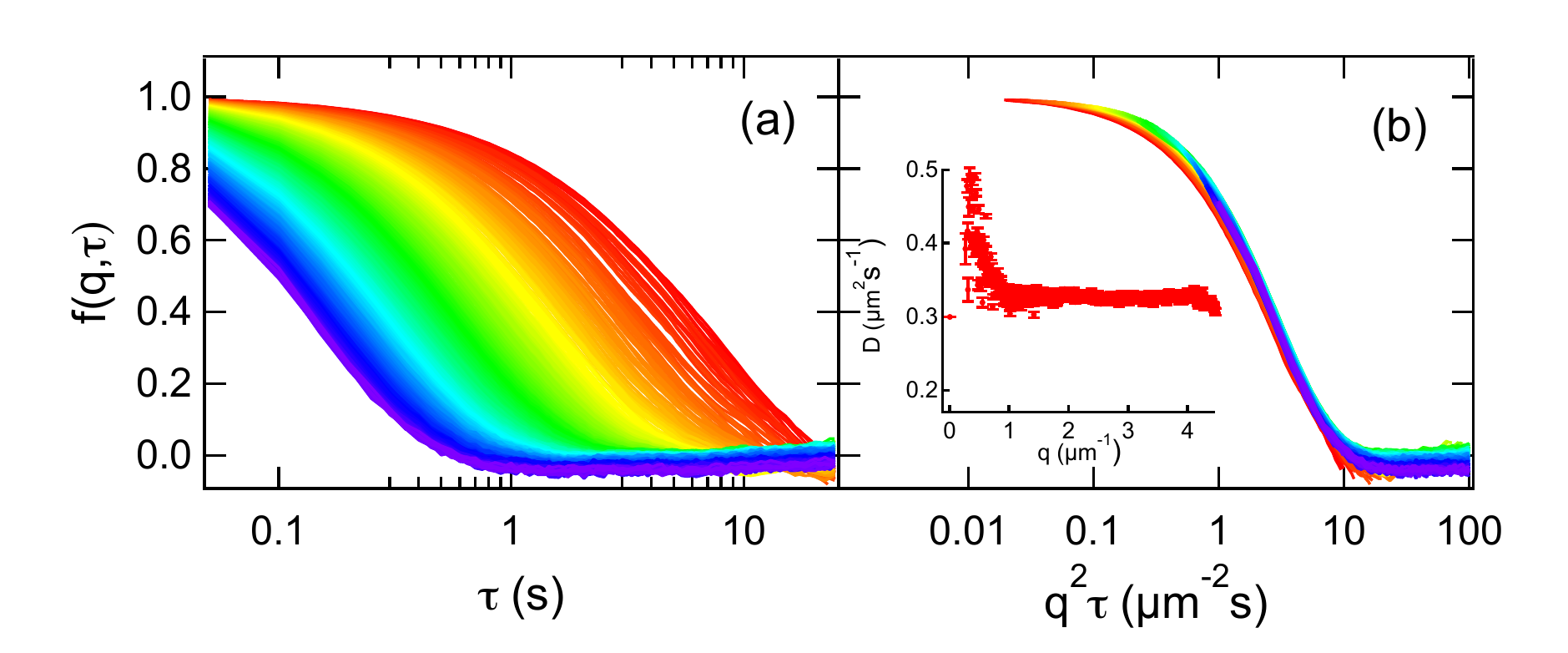}
	\end{center}
	\caption{\small{The ISF from fluorescence DDM of a sample at $\phi = 0.1\%$ with WT cells and motA mutants at a number ratio of 7:3 plotted against (a) $\tau$  and (b) $q^{2}\tau$. In each case, $q$ increases from red to purple in a rainbow scale in the range $0.62 < q < 4.46 \mu m^{-1}$.  The inset to (b) plots the fitted diffusivity of the non-motile motA (N2) cells as a function of $q$. }}
	\label{Fig_isf}
\end{figure}

\begin{figure}
	\begin{center}
	\includegraphics[width=3in]{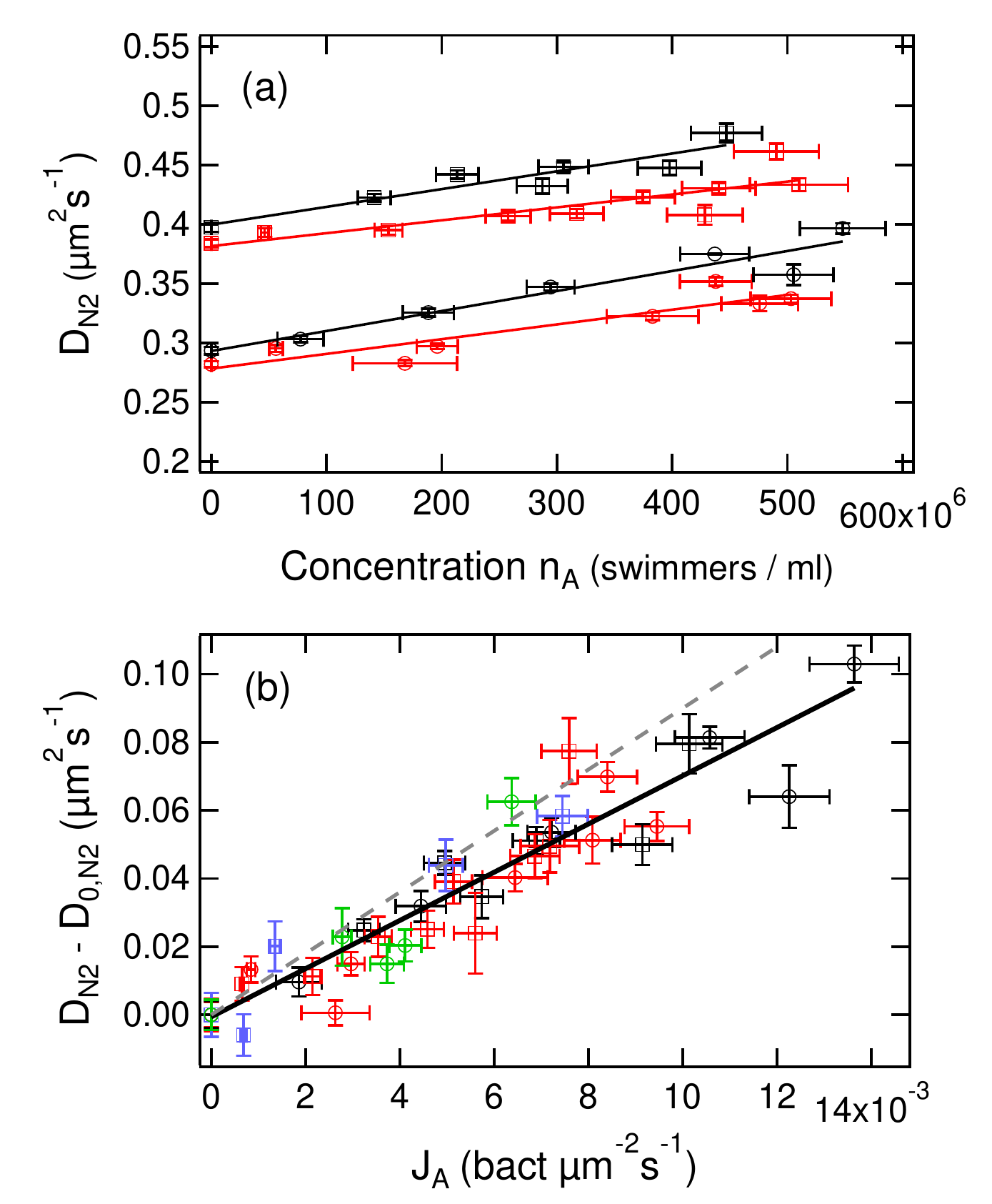} 
	\end{center}
	\caption{\small{ (a) Effective diffusivity, $D_{\rm N2}$, of motA (circles) and fliF (squares) in suspensions of {\it E. coli} AB1157 containing no glucose (red, $\bar v = 13 - 16 \mu$m/s) or glucose (black, $\bar v \lesssim 25 \mu$m/s), while $\alpha$ is varied at a fixed total cell density $\phi = 0.1\%$, with best fit lines . (b) The same data plotted as diffusivity enhancement $\Delta D = D_{\rm N2} - D_{0, \rm N2}$ for both motA and fliF data sets with and without glucose {\it versus} the active particle flux, $J_A = \bar v n_A = \bar v \alpha \phi/V$. Green and blue points: data taken by varying $\phi$ or by varying both $\phi$ and $\alpha$. Eq.~\ref{beta} fitted through all points gives $\beta = 7.1 \pm 0.4 \mu\mbox{m}^4$. Dashed line: prediction of  \cite{PushkinPreprint}. All error bars give $\pm$ standard deviation. }}
	\label{Fig:D}
\end{figure}

Figure~\ref{Fig_isf}(a) shows ISFs from fluorescence DDM performed on  a typical sample containing  70\% WT cells (M+N1) and 30\% motA  mutants (N2) at a range of $q$ values. Since only N2 cells fluoresce, the decay of these ISFs is exclusively due to the motion of the non-motile motA mutants. The data collapse against $q^{2}\tau$, Fig.~\ref{Fig_isf}(b), means that their motion is well described as diffusive, and there is little evidence for non-Gaussianity \cite{Wu2000Particle,Valeriani2011Colloids,Leptos2009Dynamics} over our experimental window. As a check, we plotted $\text{ln}(w(q,\tau))$ versus ln($\tau$), where $w(q,\tau)=-\text{ln}(f(q,\tau))/q^2$ \cite{Martinez2008,Martinez2011}. Only a hint of super-diffusion appears at very short times. The fitted values of $D_{\rm N_2}$ are shown as a function of $q$ in the inset of Fig.~\ref{Fig_isf}(b). Averaging over the flat part of $D(q)$ ($1\lesssim q \lesssim 3$ $\mu\text{m}^{-1}$) gives $\bar D^{\rm (motA)}_{\rm N_2}=0.326 \pm 0.003$ $\mu\mbox{m}^2$/s. Repeating this procedure by mixing populations of native cells and non-motile fliF or motA mutants but always at a total $\phi = 0.1\%$ yields the dependence of $D_{\rm N_2}$ on $n_A$ for each of the two different kinds of added motile cells, fliF and motA, Fig.~\ref{Fig:D}(a) (red), showing that $\Delta D$ increases linearly with $n_A$. In the same plot, we show data for swimmers in glucose with higher $\bar v$ (black). A linear dependence remains, but with a higher slope. 

Before discussing diffusion enhancement, we first comment on the thermal diffusivity of various non-motile cells. Measurements of fliF and motA mutants on their own ($J_A = 0$ in Fig.~\ref{Fig:D}(a)) gave $D^{\rm (motA)}_{0, \rm N2} = 0.29 \pm 0.01 \mu\text{m}^2\text{/s}$ and $D^{\rm (fliF)}_{0, \rm N2} =0.39 \pm 0.01\mu\text{m}^2\text{/s} \approx 1.4 \times D^{\rm (motA)}_{0, \rm N2}$. This is consistent with tracking measurements \cite{Tavaddod2011Probing}, which found that deflagellated cells diffused $\approx 50\%$ faster than cells with paralysed flagella.  The unenhanced diffusivity of native non-motile cells (N1) cannot be accessed directly, but can be obtained by performing DDM on more and more dilute suspensions of AB1157 (i.e.~using a native mixture of M + N1 cells and taking the limit $J_A \rightarrow 0$), from which we found $D_{0, \rm N1} = 0.37 \pm 0.02 \mu\text{m}^2\text{/s}$. This value is, within uncertainties, the same as that of the fliF mutants, suggesting that non-motile WT cells probably have had their flagella sheared off during preparation. Indeed, DDM measurements showed that ÔgentlerÕ preparative protocols (e.g. using blunted pipette tips to reduce shear) generally increased the motile fraction, $\alpha$.

Returning to diffusivity enhancement, we find that all four data sets in Fig.~\ref{Fig:D}(a) collapse onto a universal line if we plot the change in diffusivity, $\Delta D_{\rm N2} = D_{\rm N2} - D_{0,\rm N2}$, versus the swimmer flux, $J_A$, Fig.~\ref{Fig:D}(b). All the data in Fig.~\ref{Fig:D}(a) were obtained at fixed overall cell concentration $\phi = 0.1\%$. Figure~\ref{Fig:D}(b) includes data points in which $J_A$ had been varied by changing $\phi$ (green points) or by changing $\phi$ and $\alpha$ together (blue points). These also fit into the universal linear dependence within experimental errors. Thus, $J_A = \bar v \alpha\phi/V$ is indeed the operative variable in controlling diffusion enhancement: $\Delta D_{\rm N2} = \beta J_A$, with the best-fit value of $\beta = 7.1 \pm 0.4 \mu\mbox{m}^4$. 

An implicit assumption so far has been that the diffusivity of each non-swimmer is enhanced {\it independently}. Figure~\ref{Fig:D}(b) includes experiments performed over $0.04\% \lesssim \phi \lesssim 0.1\%$, $0 < \alpha \lesssim 0.7$, giving in each case a volume fraction of $(1-\alpha)\phi$ of non-motile cells (N1 or N1 + N2). The observed data collapse is consistent with little or no interaction between the non-swimmers. We checked this directly by measuring the diffusivity of fliF or motA cells on their own at $\phi = 0.01\%$ and $\phi = 0.1\%$, and found no change within experimental errors. 

Equation~\ref{beta} has been demonstrated before in 2D \cite{Mino2010Enhanced,Mino2012Induced}. In a bath of {\em E. coli} and $2\mu$m beads between two glass walls separated by $h=20 \mu$m, tracking gave $\beta \approx 45 \mu\mbox{m}^4$, dropping to $\approx 10 \mu\mbox{m}^4$ for $h=110 \mu$m, where bacteria and tracers remain close to one wall, so that surface effects still dominate. Our bulk value of $\beta \approx 7 \mu\mbox{m}^4$ is smaller than any of these values \cite{densities}.

Significantly, although motA and fliF have different thermal diffusivities (and therefore hydrodynamic radii), their motion is enhanced to the same extent (same $\beta$), Fig.~\ref{Fig:D}(b). Previously, enhancement in 2D close to a wall was found to be the same for $1$ and $2\mu$m tracers  \cite{Mino2010Enhanced}. These findings recall particle imaging velocimetry (PIV), where small tracers sufficiently close to being neutrally buoyant  follow the streamlines in a flow field. Corrections due to finite tracer size (radius $R$) scale as $(R/\ell)^2$ according to Fax\'en's law \cite{HappelBrenner}, for an average swimmer-tracer distance $\ell$. The `PIV regime' is obtained if $(R/\ell)^2 \ll 1$.

To estimate $\ell$, we approximate swimming {\it E. coli} cells by equivalent-volume spheres of diameter $d \approx 1.4 \mu$m, so that $\ell \sim d \phi^{-1/3} \approx 14 \mu$m at our highest total cell concentration ($\phi \approx 10^{-3}$). For native non-swimmers and fliF mutants without flagella, we take $2R = d \approx 1.4 \mu$m, so that $(R/\ell)^2 \approx 0.003$. Thus, we are in the `PIV regime' as in previous work using 1-2$\mu$m colloidal tracers \cite{Mino2010Enhanced}. However, for motA mutants with $\lesssim 10 \mu$m paralysed flagella, $(R/\ell)^2 \lesssim 0.5$. Thus, at $\phi$ somewhat higher than our highest, motA mutants will be out of the `PIV regime'; the physics in this case remains to be explored.

A tracer near a passing swimmer executes a not-quite-closed loop \cite{Yeomans2010,Childress2011,Mino2012Induced,Pushkin2013} due to far-field fluid advection, resulting in a net displacement. We adapt a theory developed for `squirmers' \cite{Childress2011} to {\it E. coli} \cite{AlexPreprint}, and show that integrating these motions over bacterial trajectories with finite persistence length accurately explains our data. 

Each flagellated {\it E. coli} cell is a pusher; the far field fluid velocity at a distance $\mathbf{r}$ from a cell is dipolar \cite{Goldstein2011}:
\begin{equation}
{\bf v}\left( {\bf r}\right) = \frac{p\,{\bf r}}{r^3} \left[ 3\cos^2{\theta} - 1\right],
\label{dipolar}
\end{equation}
with strength $p = kv$, $v$ the swimming speed and $k$ a geometric constant with dimensions (length)$^2$. We model WT cells using particles that swim straight over a persistence length $\lambda$ before randomly changing direction. 

\begin{figure}[t]
\includegraphics[width=5.5cm]{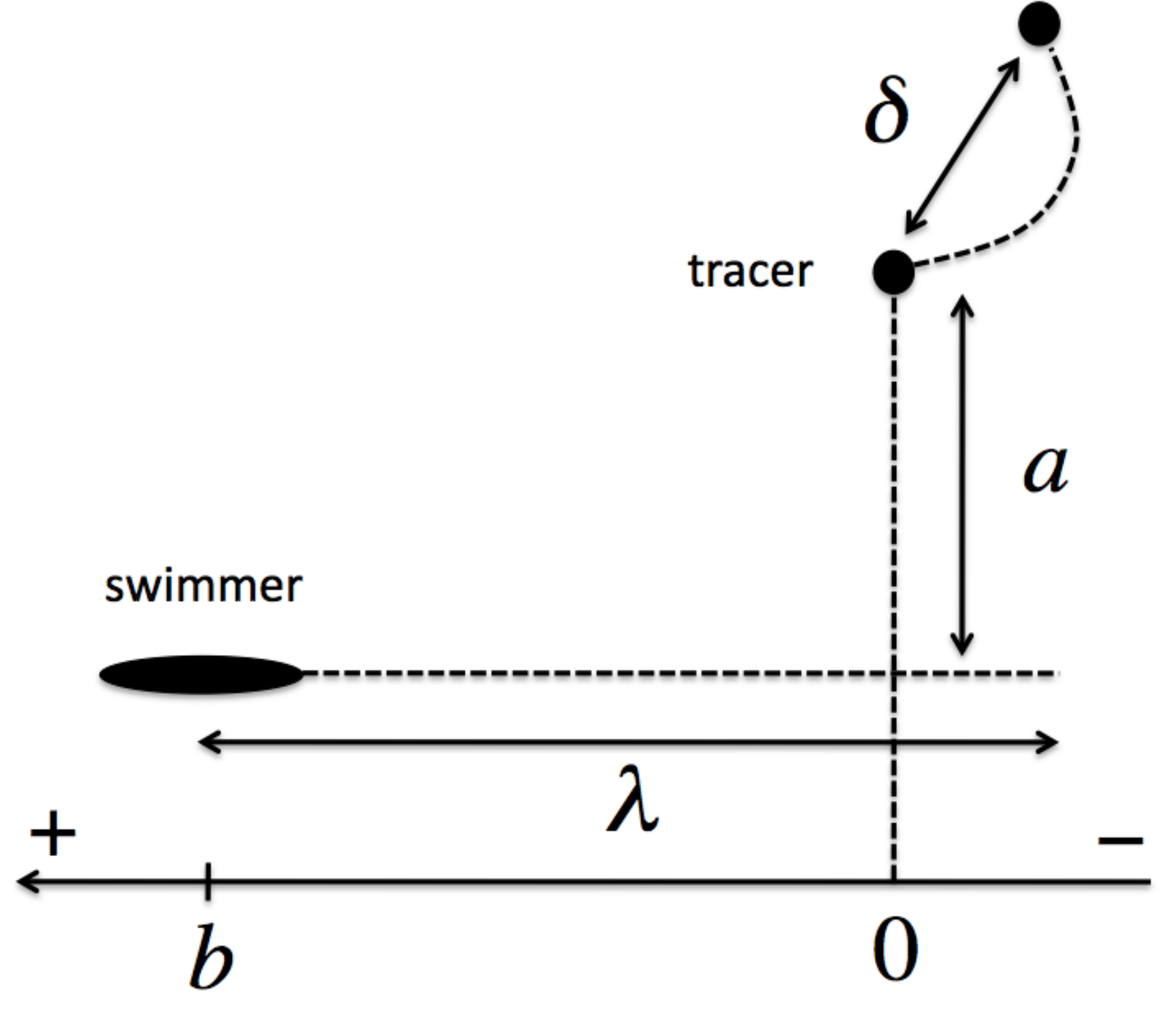}
\caption{Schematic of an `elementary scattering event' between a swimmer and a tracer. See text for definition of symbols.}
\label{event}
\end{figure}

The total displacement of a tracer is the sum of many `elementary scattering events', each of which is characterised by two `impact parameters': the distance $a$ from the tracer and the distance $b$ from the start of the straight trajectory, to the point of the closest approach, Fig.~\ref{event}. If $\lambda \rightarrow \infty$, such scattering events result in closed or almost closed loop trajectories of the tracer \cite{Yeomans2010,Childress2011,Mino2012Induced,Pushkin2013} and a slightly enhanced tracer diffusivity. Real swimmers have finite $\lambda$, so that tracers only execute parts of these loop-like trajectories, giving larger net displacements during each scattering event and higher $\Delta D$ \cite{Childress2011}. 

The mean-squared displacement of a tracer $\langle|{\bf x}(t)|^2\rangle$ comes from summing individual displacements $\delta(a,b)$ over all possible scattering configurations $a$ and $b$. Assuming identical, non-interacting, isotropic swimmers and statistically independent events \cite{Childress2011}:
\begin{eqnarray}
&&\langle|{\bf x}(t)|^2\rangle \equiv 6 \Delta D t \nonumber\\
&&\qquad \qquad = \!\left(\frac{2\pi n_A v t}{\lambda} \right)\int_{0}^{\infty} \!\!\int_{-\infty}^{\infty} \!a \delta^2(a,b) \text{d}b \text{d}a.\label{integral}
\end{eqnarray}
To understand the prefactor $(\ldots)$, note that in time $t$, each swimmer `tumbles' $v\, t/\lambda$ times to give $n_A\,v\,t/\lambda$ scattering events of the type shown in Fig.~\ref{event}. To evaluate Eq.~(\ref{integral}), we numerically integrate the tracer equations of motion, $\dot{\bf r}_t(t) = {\bf v}\left( {\bf r}_t(t) - {\bf r}_s(t)\right)$, where ${\bf r}_t(t)$ and ${\bf r}_s(t) = {\bf r}_s(0) + v\,t\,\hat{\bf e}$ are the positions of the tracer and swimmer respectively. The initial position and swimming direction $\hat{{\bf e}}$ are set by the scattering parameters $a$ and $b$. Repeating for sets of $(a,b)$ and summing up the resulting displacements, we find an enhanced diffusivity for the dipolar pusher velocity field, Eq.~\ref{dipolar}:
\begin{equation}
\Delta D = 3.44\,n_A\,v\,\left( \frac{p}{v}\right)^2 = 3.44 k^2 J_A. \label{prediction}
\end{equation}
Detailed calculations \cite{AlexPreprint} show that, as for `squirmers' \cite{Childress2011}, the numerical prefactor in Eq.~(\ref{prediction}) is not very sensitive to the range of $(\lambda, p, v)$ relevant for swimming {\it E. coli} \cite{Goldstein2011}, for which $k = 1.45 \mu\mbox{m}^2$, and Eq.~(\ref{prediction}) predicts  $\Delta D = \beta J_A$ with $\beta = 7.24 \mu\mbox{m}^4$, in remarkably good agreement with our value, Fig.~\ref{Fig:D}, of $\beta = 7.1 \pm 0.4 \mu\mbox{m}^4$. 

Previous calculations at $\lambda \rightarrow \infty$ give $\beta = 0.48 \mu\mbox{m}^4$ \cite{Mino2012Induced}, because here, tracers execute almost-closed loops \cite{Yeomans2010}.
For finite $\lambda$, the largest contribution to the integral in Eq.~\ref{integral} comes from $b=0$ and $b=\lambda$ \cite{Childress2011}. At these scattering events, Fig.~3, a swimmer starting or finishing at the point of closest approach causes a tracer to perform approximately half of the infinite-$\lambda$ almost-closed loop, giving significantly larger total displacements. Indeed, preliminary DDM measurements using a smooth swimming mutant, which has a significantly higher $\lambda$ than a run-and-tumble swimmer, showed lower enhanced diffusion of the non-swimmers.

To summarise, we have observed that the enhanced diffusion of non-motile cells in a 3D bath of motile {\it E. coli} scales linearly with the motile cell flux, Fig.~\ref{Fig:D}. The scaling is accurately accounted for by summing tracer displacements due to far-field advection induced by individual swimmers with long but finite persistence length trajectories. Interestingly, since we first submitted this work, it has suggested \cite{PushkinPreprint} that fluid entrainment is also important, which, together with advection, give $\beta = 9 \mu\mbox{m}^4$ in 3D, a value incompatible with our observations, Fig.~2b \cite{however}. 

We have worked at $\phi \lesssim 0.1\%$, where the diffusivities of non-motile fliF and motA mutants are enhanced equally. At  higher $\phi$, this situation should change, because motA cells with paralysed flagella are then too large to be considered tracers. Separately, it should be interesting to probe concentrated systems in which the density of tracers is increased until they interact with each other.

\vspace{.2cm}
AJ, VAM, ANM and WCKP were funded by an EPSRC studentship, EU FP7-PEOPLE (PIIF-GA-2010-276190), EPSRC EP/I004262/1 and EPSRC EP/J007404/1 respectively. We thank G. Dorken for assisting with plasmid transformations and M.~E. Cates, E.~Cl\'ement, G.~Mi\~{n}o, and D.~Pushkin for discussions.


\begin{thebibliography}{}
\bibitem{however} Note, however, that  \cite{PushkinPreprint} predicts $\beta = 7.6  \mu\mbox{m}^4$ for 3D advection alone, consistent with our work.
\end{thebibliography}
\end{document}